%% file: nipower2a.tex
\documentclass[a4paper,fleqn,usenatbib]{mnras}

\usepackage[T1]{fontenc}
\usepackage{ae,aecompl}

\usepackage{graphicx}	
\usepackage{amsmath}	
\usepackage{amssymb}	

\newcommand{\Msun}{$\mathrm{M}\odot$}

\title[Nickel for ASASSN--15lh]
{How much radioactive nickel does ASASSN--15lh require?}

\author[A. Kozyreva et al.]
{Alexandra~Kozyreva$^{1}$\thanks{E-mail: a.kozyreva@keele.ac.uk},
Raphael~Hirschi$^{1,2}$, 
\newauthor
Sergey~Blinnikov$^{2,3,4}$, Jacqueline den Hartogh$^{1}$
\\
$^{1}$Astrophysics group, Keele University, Keele, Staffordshire, ST5 5BG, UK \\
$^{2}$Kavli IPMU (WPI), University of Tokyo, Kashiwa, Chiba 277-8583, Japan\\
$^{3}$ITEP (Kurchatov Institute), Moscow, 117218, Russia \\
$^{4}$VNIIA, Moscow, 127055, Russia\\  
}

\date{Accepted XXX. Received YYY; in original form ZZZ}

\pubyear{2016}

\begin{document}
\label{firstpage}
\pagerange{\pageref{firstpage}--\pageref{lastpage}}
\maketitle

\begin{abstract}

The discovery of the most luminous supernova ASASSN--15lh triggered a shock-wave 
in the supernova community.  The three possible mechanisms proposed for
the majority of other superluminous supernovae do not produce a realistic
physical model for this particular supernova.  
In the present study we show the limiting luminosity available
from a nickel-powered pair-instability supernova.  We computed a few
exotic nickel-powered explosions with a total mass of nickel up to 1500
solar masses.  We used the hydrostatic configurations prepared with the
\texttt{GENEVA} and \texttt{MESA} codes, and the \texttt{STELLA} radiative-transfer code for
following the explosion of these models.  We show that 1500~solar masses of
radioactive nickel is needed to power a luminosity of
$2\times10^{\,45}$\,erg\,s$^{\,-1}$.  The resulting light curve
is very broad and incompatible with the shorter ASASSN--15lh time-scale.  This rules out
a nickel-powered origin of ASASSN--15lh.  In addition, we derive a simple
peak luminosity -- nickel mass relation from our data, which may serve to
estimate of nickel mass from observed peak luminosities.

\end{abstract}

\begin{keywords}
supernovae: general -- supernovae: individual: ASASSN--15lh -- stars: massive --
stars: evolution -- radiative transfer
\end{keywords}



\section[Introduction]{Introduction}
\label{sect:intro}

The supernova community is highly excited with the 
recent discovery of ASASSN--15lh, which is the most luminous supernova ever detected
\citep[][]{2016Sci...351..257D} among other superluminous supernovae (SLSNe)
\citep[see][for a review]{2012Sci...337..927G}.  
The estimated bolometric luminosity reported
is extremely high and reaches $(2.2 \pm 0.2)\times 10^{\,45}$\,erg\,s$^{\,-1}$.  
ASASSN--15lh is classified as a Type\,Ic
supernova, so that it requires hydrogen and helium free models to explain it.  

As ASASSN--15lh took place close to the host galaxy centre, one of the first possible
explanations is that it is a tidal disruption event 
\citep[TDE, see Supplementary materials in][]{2016Sci...351..257D}.  
The detected properties of ASASSN--15lh are different from usual
TDEs which generally possess hydrogen and helium \citep[][]{2016MNRAS.455.2918H}, although 
there are at least a few TDEs with a very small amount of hydrogen
\citep{2015ApJ...815L...5G,2016MNRAS.tmp...52K}.  

One of the most promising models is a rapidly rotating young
highly-magnetized neutron star (i.e., a magnetar)
powering the supernova ejecta through its dipole radiation \citep[][and references
therein]{2006Natur.442.1018M,2010ApJ...719L.204W,2010ApJ...717..245K,2015arXiv150802712M}.  
In the case of ASASSN--15lh, the magnetar requires a period of 1--2\,ms,
magnetic field of 0.3--1$\times10^{\,14}$\,G, and mass of the ejecta of 6\,\Msun{}
\citep{2016arXiv160101021B}, or 0.7\,ms, $2.5\times10^{\,13}$\,G, and
8.3\,\Msun{} \citep{2016arXiv160204865S}.  We discuss in more detail
the magnetar model in Section\,\ref{sect:discussion}.

Considering ASASSN--15lh as an inter\-action-po\-wer\-ed supernova seems irrelevant as it would 
require the collision of very massive hydrogen--helium poor shells and very high energies
\citep{2011ApJ...729L...6C,2012ApJ...757..178G,2013MNRAS.430.1402M,2015AstL...41...95B,2015arXiv151000834S}.  
Varying parameters (e.g. mass of the shell or energy) in the interaction
model allows one to construct an appropriate
fit for a given SLSN \citep{2013ApJ...773...76C}.  
However, interaction-powered events have lower photospheric velocities (several
thousand km\,s$^{\,-1}$).

From the stellar evolution point of view, there are a number of ideas as to how the star may produce 
shells or a dense wind, either hydrogen-free or hydrogen-rich. 
Either metal-free or metal-rich very massive stars undergo enhanced
mass--loss due to pulsations \citep{2010ApJ...717L..62Y,2011AstL...37..403F,2015A&A...573A..18M}.  
Luminous blue variables \citep[][]{1994PASP..106.1025H} or
very massive stars with similar activity are good candidates for producing dense 
circumstellar environments.  \citet{2013MNRAS.433.1114Y} show
that mass-loss rates reach several $10^{\,-3}$\,\Msun\,yr$^{\,-1}$ during core helium
burning phases of 300\,--\,500\,\Msun{} rotating and non-rotating models (eWNE phase).  The supernova shock
smashes into the surrounding medium, and instead of the usual shock
breakout the luminous supernova-like event is observed.  This is the promising mechanism for
Type\,IIn~SNSLe \citep{2014MNRAS.439.2917M}, although it remains a parameter--dependent model.  
Mass-loss can be enhanced by neutrino--emission from the stellar core or waves
\citet{2007ApJ...667..448M,2014A&A...564A..83M,2014ApJ...780...96S}.  
Nevertheless, there is no evidence for interaction signatures for
ASASSN--15lh from current observations \citep{2015ATel.8216....1M}.

Pulsational pair-instability supernova is another type of
interaction-powered scenario \citep{2007Natur.450..390W}.  A star with the
initial mass below pair-instability supernova (PISN) limit of 140\,\Msun{} does not
explode directly\footnote{The initial stellar mass ranges between 100\,\Msun{}
and 140\,\Msun{} (corresponding CO-core mass lies between 40\,\Msun{} and
60\,\Msun{}) for progenitors of pulsational pair-instability supernovae.} 
\citep{2015ASSL..412..199W}.  The star produces a
sequence of pulses of explosive nuclear burning caused by pair-creation
instability \citep{2016MNRAS.457..351Y}.  Eventually it explodes as a normal core-collapse supernova
followed by an interaction of supernova ejecta with the previously swept
stellar shells \citep{2015ApJ...814..108Y}.

The last possible known realistic scenario includes
the pair-instability explosion of a very
massive star at the upper limit of PISN mass-range ($\sim 260$\,\Msun{})
\citep{2009Natur.462..624G,2012Sci...337..927G}.  
\citet{2002ApJ...567..532H} predict that up to 55\,\Msun{} of radioactive
nickel is generated in this kind of explosion and powers a luminous
supernova event.  

In the present study we examine the maximum luminosity possible for
PISNe.  We also examine how much radioactive nickel is
needed for ASASSN--15lh, if it is considered to be purely nickel-powered.  
We describe our toy models in Section\,\ref{sect:method}, discuss the light curves
and derive a relation between nickel mass and resulting supernova peak
luminosity in Section\,\ref{sect:results} and \ref{sect:discussion}.  
We conclude in Section\,\ref{sect:conclusions}.

\section[Input models and light curve modelling]{Input models and light curve modelling}
\label{sect:method}

\input{models}

Our main input models are the following: h200, h200Ni55, 250M, 500M, 750M, 1000M, and
1500M.  We summarize their properties important for the current study in
Table\,\ref{table:models}.  
Our reference model is the h200 model.  Its earlier evolution was computed
with the stellar evolution code \texttt{GENEVA} \citep{2012A&A...537A.146E,2013MNRAS.433.1114Y}. 
Later evolution and the pair-instability explosion were simulated with the 
\texttt{KEPLER} code \citep{2014ApJ...797....9W}.  This PISN model produces 40\,\Msun{} of
radioactive nickel.  To inspect the limiting PISN luminosity (model h200Ni55), we use the
h200 model with \emph{artificially} increased, to 55\,\Msun{}, amount of nickel keeping the overall
structure as in the original h200 model.  We map h200 and h200Ni55 models
into the \texttt{STELLA} code just before shock breakout to follow up the post-explosion
evolution of the h200 and h200Ni55 ejecta
\citep[][]{1998ApJ...496..454B,2000ApJ...532.1132B,2006A&A...453..229B,2014A&A...565A..70K}.  

\begin{figure}
\centering
\includegraphics[width=0.5\textwidth]{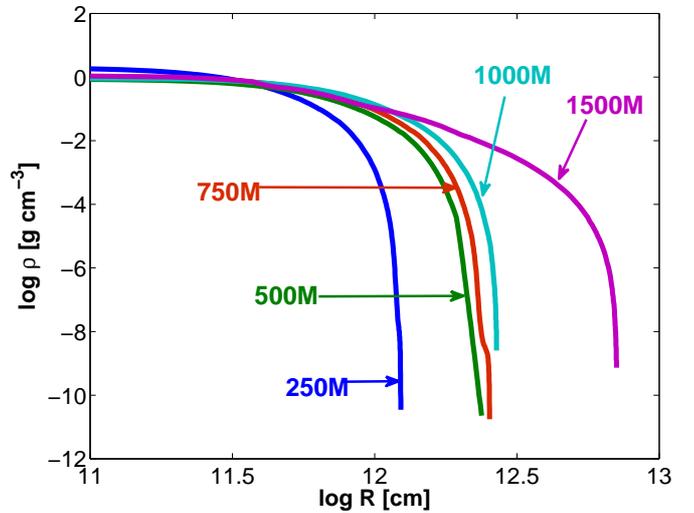}
\caption[Density structure of our very massive models.]{Density structure of
models: 250M, 500M, 750M, 1000M, and 1500M.}
\label{figure:Rhostructure}
\end{figure}

As an input for \texttt{STELLA} for the other models, we create \emph{artificial} structures in
hydrostatic equilibrium.  The hydrostatic structures were calculated with the
\texttt{GENEVA} and \texttt{MESA}\footnote{\texttt{Modules for Experiments
in Stellar Astrophysics}
http://mesa.sourceforge.net/
\citep{2011ApJS..192....3P,2013ApJS..208....4P,2015ApJS..220...15P}.} codes.  
Fig.\,\ref{figure:Rhostructure} presents the density structure 
of our input models.  For their composition, we
\emph{artificially} set pure radioactive nickel, leaving a shallow outermost layer consisting of oxygen with a total
mass not exceeding 1.7\,\Msun{}.  This is done for the numerical stability of 
the \texttt{STELLA} code.   
The explosion of these toy models in \texttt{STELLA} is done manually as an instant (lasting
0.1\,s) input of thermal energy spread out up to the surface of the
progenitor.  The explosion energy mentioned in Table\,\ref{table:models} is necessary to initiate the expansion of
the progenitor.  We chose it as: 120\,B (1\,B or Bethe = $10^{\,51}$\,erg),
500\,B, 550\,B, 700\,B, and 600\,B for 250M, 500M, 750M, 1000M, and 1500M, respectively.

\texttt{STELLA} is a one--dimensional multigroup radiation Lagrangian implicit
hydrodynamics code.  
In the present simulations we use 100 frequency bins.  The opacity is calculated
based on about 160,000 spectral lines from \citet{1995all..book.....K} and
\citet{1996ADNDT..64....1V}.  
Deposition from nickel and cobalt
decay is treated in a one-group approximation according to \citet{1995ApJ...446..766S}.

\section{Results}
\label{sect:results}

\begin{figure}
\centering
\includegraphics[width=0.5\textwidth]{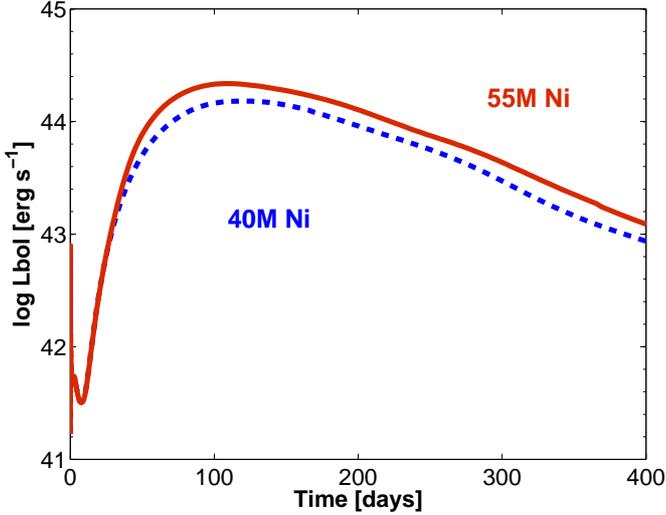}
\caption[Bolometric light of the model h200 with
40\,\Msun{} and h200Ni55 with 55\,\Msun{} of nickel.]{Bolometric light
curves of the PISN model h200 with 40\,\Msun{} of nickel and
h200Ni55 with 55\,\Msun{} of nickel.  h200Ni55 demonstrates the PISN limiting luminosity.}
\label{figure:PISNlimit}
\end{figure}

We show the comparison between our reference model h200 and the model h200Ni55
in Fig.\,\ref{figure:PISNlimit}.  The peak luminosity of the
nickel-powered light curve is expected to depend on the mass of radioactive nickel 
according to the following relation \citep{1979ApJ...230L..37A}:
\begin{equation}
L_\mathrm{\,peak} \sim M_\mathrm{\,Ni} \, \varepsilon(t_\mathrm{\,peak}) \,,
\label{equation:LNI}
\end{equation}
where $\varepsilon(t_\mathrm{\,peak})$ is the decay function of nickel and cobalt.  
In Fig.\,\ref{figure:PISNlimit}, the difference in the peak luminosity between models with 40\,\Msun{} and
55\,\Msun{} of nickel is exactly what Equation\,\ref{equation:LNI} predicts and is equal to 0.16\,dex.  
We  chose 55\,\Msun{} of nickel for the h200Ni55 model, as it
represents the upper
luminosity limit which a PISN is able to produce \citep{2002ApJ...567..532H}.  
Thus, the PISN limit is: peak bolometric luminosity equal to
$2.2\times10^{\,44}$\,erg\,s$^{\,-1}$,
or a bolometric magnitude of $-22.12$\,mag.  

\begin{figure}
\centering
\includegraphics[width=0.5\textwidth]{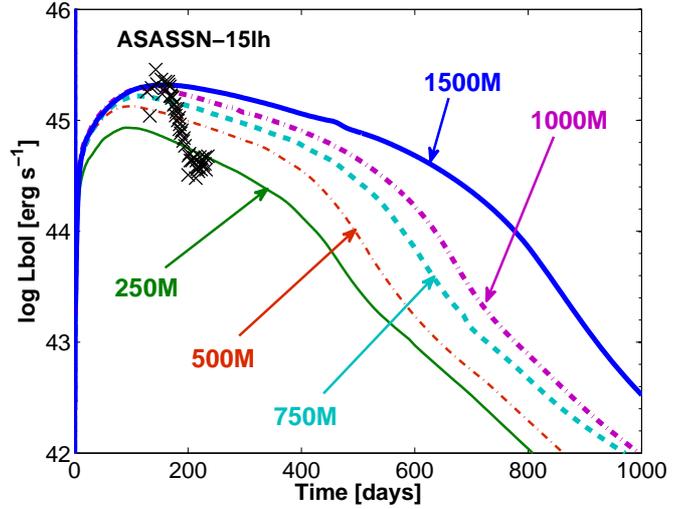}
\caption[Bolometric light curve of our nickel-rich models]
{Bolometric light curves of our nickel-rich models: 250M (thin solid), 500M
(thin dash-dotted), 750M (thick dashed), 1000M (thick dash-dotted), and 1500M
(thick solid).  Crosses represent estimated
bolometric luminosity of ASASSN--15lh 
\citep{2016Sci...351..257D}.}
\label{figure:Ni750}
\end{figure}

Fig.\,\ref{figure:Ni750} shows the results for our exotic explosions of
progenitors with the following amounts of nickel: $\sim$250, 500, 750,
1000, and 1500\,\Msun{}.   
Models reach the following peak luminosities: $8.6\times10^{\,44}$,
$1.3\times10^{\,45}$, $1.7\times10^{\,45}$,
$1.9\times10^{\,45}$, and $2.1\times10^{\,45}$\,erg\,s$^{\,-1}$,
respectively.  
According to Equation\,\ref{equation:LNI}, 619\,\Msun{} of nickel supports the
peak of ASASSN--15lh.  However, we found that our exotic explosions 
decline from Equation\,\ref{equation:LNI} 
displaying a weaker peak.  This happens because nickel is spread out
up to the edge of the star, and a large fraction of gamma-photons escape freely without
interaction with the ejecta material \citep{1979ApJ...230L..37A}.  Hence, the ``effective nickel mass'' is lower.  
Fig.\,2 from \citet{2015MNRAS.454.4357K} also shows that a gradually truncated input model
provides a slightly weaker peak even without changing nickel mass.  

\begin{figure}
\centering
\includegraphics[width=0.5\textwidth]{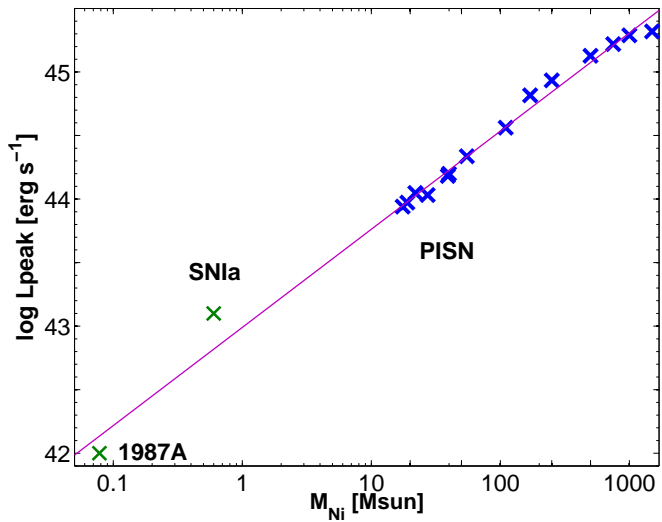}
\caption[Luminosity\,--\,nickel mass relation]
{Luminosity\,--\,nickel mass relation in the nickel mass range
0.05\,--\,1500\,\Msun{}.  The thin line indicates the least-squared linear
fit.  SN\,1987A and basic SN\,Ia are superposed.}
\label{figure:LMNi}
\end{figure}

Fig.\,\ref{figure:LMNi} illustrates the relation between $L_\mathrm{peak}$ and
$M_{\mathrm{\,Ni}}$
based on results of the current study and other high-mass PISN
models\footnote{Proper PISN explosions: P200,
P250 (Ni 18\,\Msun{}, 27\,\Msun{}, Kozyreva\,et\,al., in preparation), 250M
\citep[Ni 19\,\Msun{},][]{2014A&A...565A..70K}, ``260'' \citep[Ni
22\,\Msun{},][]{2015ApJ...799...18C}, h200
\citep[Ni 39\,\Msun{},][]{2014ApJ...797....9W}, he130 \citep[Ni 40\,\Msun{},][]{2011ApJ...734..102K}. 
Artificial explosions with artificially increased nickel mass: 55\,\Msun{} based on h200,
110\,\Msun{} and 170\,\Msun{} based on 250M.}.  
By fitting our data with 
a linear least-squared fit in Fig.\,\ref{figure:LMNi}, we obtain the following relation:
\begin{equation}
\log L_{\mathrm{\,peak}} = 42.99\, (\pm 0.12) + 0.77\, (\pm 0.05) \, \log M_{\mathrm{\,Ni}} \,.
\label{LMNia}
\end{equation}
The relation may be used to estimate the nickel mass needed for
a given peak bolometric luminosity, assuming it is a nickel-powered light
curve.  
In addition, SN\,1987A \citep{1989ARA&A..27..629A} and
basic SN\,Ia \citep[W7 model,][]{1984ApJ...286..644N} are included in
Fig.\,\ref{figure:LMNi}.  
The relation given in Eq.\,\ref{LMNia} resembles that for SNe\,Ia
\citep{2005A&A...431..423S,2006A&A...453..229B,2006A&A...460..793S}.  

Based on our simulations, 
the bolometric luminosity of
ASASSN--15lh requires an explosion in which at least 1000\,\Msun{}
of radioactive nickel is produced.  It is obvious that there are no known
stellar explosions which are capable of generating this huge nickel mass.  We
discuss a few aspects of producing high nickel mass in Section\,\ref{sect:discussion}.

A high fraction of nickel results not only in the high peak luminosity, but
also in slow evolution of a light curve.  The light curve for the 1500M model
lasts about 1000\,d above $10^{\,43}$\,erg\,s$^{\,-1}$.  This fact resembles the well-known
Pskovskii--Phillips relation for SN\,Ia 
\citep{1977SvA....21..675P,1993ApJ...413L.105P}, i.e., higher mass of nickel
supports brighter peak and broader light curve.

\section{Discussion}
\label{sect:discussion}

In this section we discuss the maximum luminosity for PISN explosions.  
The grid of models from \citet{2002ApJ...567..532H} shows the results of
the simulation of non-rotating
zero-metallicity models in the helium-core mass range between 65\,\Msun{} and
135\,\Msun{}.  The initial corresponding main-sequence star mass ranges
from 140\,\Msun{} to 260\,\Msun{}.  However, the evolutionary parameter space
greatly exceeds this grid of very massive stars \citep{2015ASSL..412..199W}.

There is no difficulty to produce stars as massive
as a thousand solar masses, although they were not observed yet
\citep[][]{1969ARA&A...7..553W,2002Sci...295...93A,2008A&A...477..223Y,2009Ap&SS.324..271V,2009ApJ...703.1810H,2013ApJ...778..178H}.  
Such large stellar objects may be formed in a two-step process
through accretion on the already formed proto-star \citep{2011ApJ...732...20K}. 
However, the uncertainty in mass-loss rates for very massive stars implies
large uncertainties in their evolution \citep{2015ASSL..412...77V}.

The crucial aspect in the pair-instability explosion is the balance between binding
energy of the star and nuclear energy released in the explosive oxygen and silicon burning.  
Binding energy is proportional to its gravitational energy (with a
factor of $\gamma-{4\over3}$), i.e. its mass.  
For higher massive objects, density is lower,
because of radiation-dominated pressure, which leads to larger radii
(Fig.\,\ref{figure:Rhostructure} in this {\it Letter}, and
\citealt{1969ARA&A...7..553W}).  
A higher mass object is more ``spongy'', and its specific binding
energy ($\varepsilon/M$) is lower \citep[see Table\,3 in][]{1967SvA....10..604B},  
therefore, it is easy to unbind such systems.  
Roughly, half of the mass of the star is left after wind mass-loss, and it becomes
a carbon-oxygen core 
\citep{1982sscr.conf..303B}.  
Once the carbon-oxygen core undergoes the pair-creation instability, oxygen
burns and provides the essential fraction of explosive energy, which
disrupts the star.  The released nuclear energy then depends linearly on
the mass of the carbon-oxygen core.  
Hence, the upper-mass boundary for the
pair-instability explosion might be higher than 260\,\Msun{}.  
It is, however, not expected to reach anywhere near a thousand solar masses
of radioactive nickel.

Rotation plays a role in reducing the gravitational force, but not drastically
\citep{1974ApJ...189..535F}.  If a carbon-oxygen core retains large angular momentum
\citep{1985A&A...149..413G}, the PISN upper-mass boundary may
slightly increase.

An object as massive as 500--1000\,\Msun{} or higher is
a subject of general relativity.  According to general relativistic effects,
these supermassive objects in general collapse into black holes, and nothing 
prevents contraction.  
However, some supermassive star models in a narrow mass-range around
55,000\,\Msun{} explode (see studies by \citet{1973ApJ...183..941F},
\citet{1986ApJ...307..675F}, \citet{2013ApJ...775..107J}, \citet{2013ApJ...777...99W},
\citet{2013ApJ...778...17W}, and \citet{2014ApJ...790..162C} which include
general relativity), but they do not produce large nickel masses.  
Adding these factors together cannot help in producing
1000\,\Msun{} of nickel.  Therefore, we conclude that ASASSN--15lh is clearly not a
pair-instability explosion.

As we mention in Section\,\ref{sect:intro}, if consider a magnetar
origin for ASASSN--15lh, the magnetar requires a period of 1--2\,ms,
magnetic field of 0.3--1$\times10^{\,14}$\,G, and mass of the ejecta of 6\,\Msun{}
\citep{2016arXiv160101021B}, or 0.7\,ms, $2.5\times10^{\,13}$\,G, and
8.3\,\Msun{} \citep{2016arXiv160204865S}.  
However, it is not very clear whether a magnetar may store such a high rotational energy
($3\times10^{\,52}$\,erg)  and strong magnetic field \citep{2015arXiv150802712M}.  
Independent detailed simulations show that if a magnetar powers a supernova ejecta, the
resulting event is not luminous (Badjin\,et\,al.\,in preparation).  An
ambiguous aspect is how the magnetic dipole radiation (a Poynting flux) is converted
into thermal energy of the ejecta, as high-energy photons produce
electron-positron pairs and, in turn, high gamma-ray opacity \citep[see][for
discussion]{2015arXiv150703645K}.  This
cascading process supplies an uncertainty to the thermalization timescale and
thermalization efficiency, and prevents a high photon flux, hence, the production 
of a luminous supernova.  Nevertheless, it may be that independent of our
understanding of the microphysics, dipole radiation indeed is converted into
ejecta thermal energy, and the magnetar-powered mechanism remains the best
candidate for such luminous events.


\section{Conclusions}
\label{sect:conclusions}

In the present study we carried out post-explosion evolution of a PISN with
a total mass of nickel equal to 55\,\Msun{}.  55\,\Msun{} of nickel is roughly the upper
published limit available in a pair-instability explosion.  The limiting bolometric luminosity
of a PISN corresponds to $2.2\times10^{\,44}$\,erg\,s$^{\,-1}$, i.e.,
--22.12\,mag.  We show that the peak luminosity of ASASSN--15lh 
significantly exceeds the upper threshold available for PISN explosions, therefore, it is
definitely not a PISN.

On top of that we carried out simulations of a grid of exotic explosions with total mass of
nickel: 250, 500, 750, 1000, and 1500\,\Msun{}.  
In the assumption of purely nickel-powered light curves the peak luminosity of
$2\times10^{\,45}$\,erg\,s$^{\,-1}$ requires at least 1000\,\Msun{} of nickel. 
However, the resulting light curve is too broad (1000\,d) and incompatible
with the shorter time-scale of ASASSN--15lh.

The late-time observations of ASASSN--15lh will shed light on the origin of this unique
event.  One of the possibilities is that this is a tidal disruption
event in which a companion star firstly loses entirely its hydrogen and helium layers at
an earlier time and then accretes on to a massive black hole.  Another possible
scenario includes a fast rotating highly-magnetized neutron star, although the
required rotation and magnetic fields are high.

\section*{Acknowledgements}

The authors acknowledge support from EU-FP7-ERC-2012-St grant\,306901.  The
authors thank Alexander Heger for very useful comments which helped to
improve the manuscript.  AK thanks Andrea Cristini for proofreading the manuscript.  
SB is supported by a grant from the Russian Science Foundation (project
number 14-12-00203).

\addcontentsline{toc}{section}{Acknowledgements}

\bibliographystyle{mnras}

\bsp	
\label{lastpage}
\end{document}

%% file: models.tex
\begin{table}
\caption[PISN models]
{Characteristics of PISN models mapped into \texttt{STELLA}.  
Final mass M$_\mathrm{fin}$ and Ni mass are in solar masses.  
The energy unit is Bethe (B) which is $10^{\,51}$~erg.}
\label{table:models}
\begin{center}
\begin{tabular}{r|c|c|c}
model  & M$_\mathrm{fin}$ & Ni        & E$_\mathrm{expl}$ \\
name   & [\Msun{}]        & [\Msun{}] & [B]               \\
\hline
h200   & 131 &  40   & 87        \\
h200Ni55& 131 & 55   & 87        \\
250M   & 250 & 249  & 120        \\ 
500M   & 499.9 & 498.2  & 500        \\ 
750M   & 749.5 & 748.3  & 550        \\
1000M  & 1000& 998.8 & 700        \\
1500M  & 1500& 1499 & 600        \\
\end{tabular}
\end{center}
\end{table}